\documentclass[12pt,a4paper]{article}

\usepackage{amsmath,amssymb}

\newcommand{\bsl}{\boldsymbol}
\newcommand{\rcm}{\mathbf{R}_{\text{cm}}}

\begin{document}

\author{I.M.Narodetskii, M.A.Trusov \\
{\itshape Institute of Theoretical and Experimental Physics,
Moscow, Russia}}

\title{The heavy baryons in the nonperturbative string approach}

\maketitle

\large

\begin{abstract} \noindent
We present some piloting calculations of the short-range
correlation coefficients for the light and heavy baryons and
masses of the doubly heavy baryons $\Xi_{QQ'}$ and $\Omega_{QQ'}$
($Q,Q'=c,b$) in the framework of the simple approximation within
the nonperturbative QCD approach.
\end{abstract}

\newpage

\section{Introduction}

The observation of $B_c^+$ meson by the CDF collaboration
\cite{CDF} opens a new direction in the physics of hadrons
containing two heavy quarks. Presently at the LHC, $B$-factories,
and the Tevatron with high luminosity, several experiments have
been proposed, in which there is a possibility to identify and
study hadrons containing two heavy quarks, like doubly-charm
baryons $(ccq)$ or baryons $(bcq)$ with charm and beauty\footnote{
Here, and throughout this paper, $q$ denotes a light quark $u$ or
$d$.}. In the more distant future the next generation experiments
with high bottom quark production rate will provide excellent
possibilities for the study bottom baryons and their decays. In
view of this project, it is important to have safe theoretical
predictions for heavy baryon masses as a guide to the experimental
search of these hadrons.

A number of authors \cite{CI86}-\cite{LiOn} have already
considered baryons containing two heavy quarks in anticipation of
future experiments which may discover these particles. In most of
these works, however, theoretical predictions are somewhat biased
by the introduction of the additional dynamical assumptions and
supplementary dynamical parameters like constituent quark masses
in addition to the only one parameter really pertinent to QCD --
the overall scale of the theory $\Lambda_{\text{QCD}}$.

The purpose of this paper is to calculate the masses of the heavy
baryons in a simple approximation within the nonperturbative QCD,
developed in \cite{Si88}-\cite{DKS94}. This method has been
already applied to study baryon Regge trajectories \cite{FS91}
and, very recently, for computation of magnetic moments of light
baryons \cite{KS00}. The essential point of this paper is that it
is very reasonable that the same method should also hold for
hadrons containing heavy quarks. In this work  we will concentrate
on the masses of doubly heavy baryons. As in \cite{KS00} we take
as the universal parameter the QCD string tension $\sigma$, fixed
in experiment by the meson  and baryon  Regge slopes. We also
include the perturbative Coulomb interaction with the frozen
coupling $\alpha_s(\text{1~GeV})=0.39$. The basic feature of the
considered approach is the dynamical calculation of  the quark
constituent masses $m_i$ in terms of the quark current masses
$m^{(0)}_i$. This is done using the einbein (auxiliary fields)
formalism, which is proven to be rather accurate in various
calculations for relativistic systems. The einbeins are treated as
the variational parameters which are to be found form the
condition of the minimum of baryon eigen energies \cite{KaNeSi}.

\section{Formalism}

The starting point of the approach is the Feynman--Schwinger
representation of the $3q$ Green's function, where the role of
"time"  parameter along a quark path is played by the
Fock--Schwinger proper time. The final step is the derivation of
the c.m.  Effective Hamiltonian (EH) containing the dynamical
quark masses as parameters. For many details see the original
papers \cite{Si88}-\cite{DKS94}.

Consider a baryon consisting of three quarks with arbitrary masses
$m_i$, $i=1,2,3$. In what follows we confine ourselves to
consideration of the ground state baryons without radial and
orbital excitations in which case tensor and spin-orbit forces do
not contribute perturbatively. Then only the spin-spin interaction
survives in the perturbative approximation. The EH has the
following form:
\begin{equation}
\label{EH}
H=\sum\limits_{i=1}^3\left(\frac{m_i^{(0)2}}{2m_i}+
\frac{m_i}{2}\right)+H_0+V,
\end{equation}
where $m_i^{(0)}$ are the current quark masses and $m_i$ are the
dynamical quark masses to be found from the minimum condition (see
Eq. (\ref{11}) below). Since $m_i\gg m_i^{(0)}$ for light quarks,
but $m_i\sim m_i^{(0)}$ for heavy quarks, each light quark
contributes to the baryon mass an additional mass $\sim m_i/2$
(not $m_i$ as in the ordinary nonrelativistic quark model),
whereas each heavy quark contributes an additional mass $\sim
m_i$. The dynamical quark masses are evaluated from the equations
defining the stationary points of the baryon mass $M_B$ as
function of $m_i$
\begin{equation}
\frac{\partial M_B(m_i)}{\partial m_i}=0\, . \label{11}
\end{equation}
Let $\bsl{r}_i$ be the quark coordinates. The kinetic momentum
operator $H_0$ in Eq. (\ref{EH}) acquires the familiar form
\begin{equation}
\label{H_0}
H_0=-\frac{1}{2m_1}\frac{\partial^2}{\partial\bsl{r}_1^2}
-\frac{1}{2m_2}\frac{\partial^2}{\partial\bsl{r}_2^2}
-\frac{1}{2m_3}\frac{\partial^2}{\partial\bsl{r}_3^2}.
\end{equation}
$V$ is the sum of the perturbative Coulomb-like one gluon exchange
potential and the string potential. The Coulomb-like potential is
\begin{equation}
\label{coulomb}
V_c=-\frac{2\alpha_s}{3}\sum\limits_{i<j}\frac{1}{|\bsl{r}_{ij}|},
\end{equation}
where the factor $2/3$ is the value of the quadratic Casimir
operator for the group $SU_c(3)$. The string potential has been
calculated in \cite{FS91} as the static energy of the three heavy
quarks
\begin{equation}
\label{string} V_{\text{string}}(\bsl{r}_1,\bsl{r}_2,
\bsl{r}_3)=\sigma R_{\text{min}},
\end{equation}
where $R_{\text{min}}$ is the sum of the three distances
$|\bsl{r}_i|$ from the string junction point, which for simplicity
is chosen as coinciding with the centre-of-mass coordinate $\rcm$.

\section{Hyper Radial Approximation}

We use the hyperspherical formalism approach (for detail see
original papers \cite{Si66}). In the hyperradial approximation
(HRA) corresponding to the truncation of the wave function
$\psi(\{\bsl{r}_i\})$ by the component with grand orbital momentum
$K=0$ the three-quark wave function depends only on the
hyperradius $R^2=\bsl{\rho}^2+\bsl{\lambda}^2$, where $\bsl{\rho}$
and $\bsl{\lambda}$ are the three-body Jacobi
variables\footnote{for their definition see Appendix}, and does
not depend on angular variables. The confining potential
(\ref{string}) has a specific three-body character. However, this
potential as well as the Coulomb potential in Eq. (\ref{coulomb})
is smooth in the sense that the HRA (where only the part of the
potential which is invariant under rotation in the six-dimensional
space spanned by the Jacobi coordinates is taken into account) is
already an excellent approximation. The HRA neglects the mixed
symmetry components of the three-quark wave function, which appear
in the higher approximations of the hyperspherical formalism
\cite{Si66}. Introducing the reduced function\footnote{In what
follows we omit the value of $K=0$ to avoid subscripts. Note that
the radially symmetric component with $K=0$ is the dominant one in
the three-quark wave function.} $\chi(R)=R^{5/2}\psi(R)$ and
averaging $V= V_c+ V_{\text{string}}$ over the six-dimensional
sphere one obtains the Schr\"odinger equation
\begin{equation} \label{shr}
\frac{d^2\chi(R)}{dR^2}+2\mu\left[E-W(R)-\frac{15}{8\mu
R^2}\right]\chi(R)=0, \end{equation} where $\mu$ is an arbitrary
parameter with the dimension of mass which drops off in the final
expressions. The last term in (\ref{shr}) represents the
three-body centrifugal barrier and $W(R)$ is the average of the
three-quark potentials over the six-dimensional sphere:
\begin{equation} \label{W} W(R)=\langle V\rangle=-\frac{a}{R}+bR\,
, \end{equation} with
\begin{equation} \label{ab} a=\frac{2\alpha_s}{3}\cdot
\frac{16}{3\pi}\sum\limits_{i<j}\alpha_{ij},~~~
b=\sigma\cdot\frac{32}{15\pi}\sum\limits_{i,j}\gamma_{ij}.
\end{equation} The mass depending constants $\alpha_{ij}$ and
$\gamma_{ij}$ are defined by Eqs. (\ref{alpha}) and (\ref{gamma})
in the Appendix.

It is convenient to introduce a new variable $x=R\sqrt{\mu}$, to
eliminate an artificial dependence of Eq. (\ref{shr}) on $\mu$,
then the equation (\ref{shr}) becomes \begin{equation} \label{eqx}
\chi''(x)+2\left(E-U(x)-\frac{15}{8x^2}\right)\chi(x)=0\, ,
\end{equation} where
\begin{equation} \label{U(x)}
U(x)=-\frac{a\sqrt{\mu}}{x}+\frac{b}{\sqrt{\mu}}x\, .
\end{equation} Since $a\sim 1/\sqrt{\mu}$, $b\sim\sqrt{\mu}$ (see
Eqs. (\ref{alpha}), (\ref{gamma})), the eigenvalue $E$ in
(\ref{shr}) does not depend on $\mu$.

\section{The quark dynamical masses}

Equation (\ref{eqx})  applied to the nucleon $(m_q^{(0)}\sim 0)$
yields the dynamical mass $m_q$ of the light quark, and applied to
the strange hyperons gives the strange quark mass $m_s$. In the
same manner application of this equation to the charm and beauty
baryons yields the constituent masses of $c$- and $b$-quarks. In
our calculations we use the same parameters as in \cite{KN00},
namely $\sigma=0.17$~GeV, $\alpha_s=0.4$, $m^{(0)}_q=0.009$ GeV,
$m^{(0)}_s=0.17$ GeV, $m^{(0)}_c=1.4$ GeV, and $m^{(0)}_b=4.8$
GeV.

We solve Eq. (\ref{eqx}) using both the quasiclassical and
variational solutions. The first approach is based on the
well-known fact that interplay between the centrifugal term and
the confining potential produces a minimum of the effective
potential specific for the three-body problem. The numerical
solution of (\ref{eqx}) for the ground state eigen energy may be
reproduced on a per cent level of accuracy by using the parabolic
approximation for the effective potential \cite{KNS87},
\cite{KPS90}. This approximation provides an analytical expression
for the eigen energy. The potential $\tilde
U(x)=U(x)+\dfrac{15}{8x^2}$ has the minimum at a point $x=x_0$,
which is defined by the condition $\tilde{U}'(x_0)=0$, {\it i.e.}:
\begin{equation} \label{x_0} \frac{b}{\sqrt{\mu}}
x_0^3+(a\sqrt{\mu}) x_0-15/4=0\, . \end{equation}
Expanding
$\tilde U(x)$ in the vicinity of $x=x_0$ one obtains:
\[
\tilde U(x)\approx \tilde U(x_0)+\frac{1}{2}\tilde
U''(x_0)(x-x_0)^2,
\]
{\it i.e.} the potential of the harmonic oscillator with the
frequency $\omega=\sqrt{\tilde U''(x_0)}$. Therefore the energy
eigenvalue is \begin{equation} \label{eigenvalue} E_0\approx
\tilde U(x_0)+\frac{1}{2}\omega\, .
\end{equation}

In Table 1 we show the dynamical masses $m_i$ and the ground state
eigenvalues $E_0$ for various baryons calculated using the
procedure described above. Our values of light quark mass $m_q$
qualitatively agree with the results of \cite{KN00} obtained from
the analysis of the heavy-light ground meson states, but $\sim 60$
MeV higher than those of \cite{FS91}, \cite{KS00}. This difference
is due to the different treatment of the Coulomb and spin-spin
interactions. In \cite{FS91} both interactions have not been
included and the light quark mass has been calculated from the fit
of the mass of $\Delta(1232)$ where the Coulomb-like potential and
the spin-spin interaction seem to balance each other. In
\cite{KS00} the smeared spin-spin interaction for the light quarks
has been included into Eq. (\ref{11}) defining the dynamical mass
of the light quark. In our calculation as in \cite{KN00} we
include the Coulomb-like term, but neglect the spin-spin
interaction.

There is no good theoretical reason why dynamical quark masses
need to be the same in different mesons and baryons. From the
results of Table 1 we conclude that the dynamical masses of the
light quarks ($u$, $d$, or $s$) are increased by $\sim 100-150
\text{ MeV}$ when going from the light to heavy baryons. For the
heavy quarks ($c$ and $b$) the variation in the values of their
dynamical masses is marginal. In Table 2 we compare the quark
masses in $\Lambda_Q$ and $\Xi_Q$ baryons with those calculated in
\cite{KN00} in $D$ and $B$ mesons. One observes that the masses of
the light quarks in baryons are slightly smaller than those in the
mesons. The small variations in the values of $m_c$ and $m_b$ are
within the accuracy of our calculations.

\section{Correlation functions for the baryons}

For many applications the quantities $\langle\psi|
\delta^{(3)}(\bsl{r}_j-\bsl{r}_i)|\psi\rangle$ are needed. To
estimate effects related to the baryon wave function we solve Eq.
(\ref{eqx}) by the variational method. We introduce a simple
variational ans\"atz for $\chi(x)$
\begin{equation} \label{trial}
\chi(x)=2\sqrt{2}p^3x^{5/2}e^{-p^2x^2}\, , \end{equation} where
$p$ is the variational parameter, and the numerical factor is
chosen so that $\int\chi^2(x)dx=1$. The trial three-quark
Hamiltonian admits explicit solutions for the energy, the wave
function, and the density matrix: \begin{equation}
E_0\approx\min\limits_p E(p), \end{equation} where
\begin{equation} E(p)=\langle\chi|H|\chi\rangle=
3p^2-(a\sqrt{\mu})\cdot\frac{3}{4}\sqrt{\frac{\pi}{2}}\cdot
p+(b/\sqrt{\mu})\cdot\frac{15}{16}\sqrt{\frac{\pi}{2}}\cdot
p^{-1}. \end{equation}

The density matrix (the correlation function)
$f_{ijk}(\bsl{r}_{ij})$ in a baryon \{ijk\} is defined as:
\begin{equation}
f_{ijk}(\bsl{r}_{ij})=\alpha_{ij}^3\int|\psi(\alpha_{ij}\bsl{r}_{ij},\bsl{\lambda}_{ij})|^2d^3\bsl{\lambda}_{ij}
\end{equation}
so that
\begin{equation}
\int f_{ijk}(\bsl{r}_{ij})d^3\bsl{r}_{ij}=
\iint|\psi(\bsl{\rho}_{ij},\bsl{\lambda}_{ij})|^2d^3\bsl{\lambda}_{ij}d^3\bsl{\rho}_{ij}=1\,
.
\end{equation}

For the trial function (\ref{trial}) $f_{ijk}(\bsl{r}_{ij})$ are
evaluated explicitly:
\begin{equation}
f_{ijk}(\bsl{r}_{ij})=\left(\frac{\xi_{ij}}{\pi}\right)^{3/2}
e^{-\xi_{ij}|\bsl{r}_{ij}|^2}\, ,
\end{equation}
with
\begin{equation}
\xi_{ij}=2p_0^2\cdot\mu_{ij},
\end{equation}
where $\mu_{ij}$ is the reduced mass of the quarks $i$ and $j$,
and $p_0$ is to be found from the condition
\[
\left.\frac{dE}{dp}\right|_{p=p_0}=0\, .
\]
The expectation values $f_{ijk}(\bsl{r}_{ij})$ depend on the third
or `spectator' quark through the three-quark wave function.

Let us define the quantities \begin{equation} \label{R_{ijk}}
R_{ijk}=f_{ijk}(0)=\left(\frac{\xi_{ij}}{\pi}\right)^{3/2}
\end{equation}
The corresponding quantity for a meson is denoted as $R_{ij}$. The
results of the variational calculations are given in Table 3 where
for each baryon we show the variational parameters $p_0$, the
quantities $R_{ijk}$ (in units of GeV${}^3$), and the average
distances $\bar{r}_{ij}=\sqrt{\langle\bsl{r}_{ij}^2\rangle}$ (in
units of fm). The variational estimations of $E_0$ and quark
dynamical masses do not differ from those shown in Table 1.

Comparing the results of Table 3 with those of \cite{KN00} we
obtain (see Table 4)\footnote{ Inequalities (\ref{Li_1}) and
(\ref{Li_2}) were first suggested in \cite{Li96} from the observed
mass splitting in mesons and baryons.}

\begin{equation}
\label{Li_1}
R_{ijk}~<~ \frac{1}{2}R_{ij},
\end{equation}
and
\begin{equation}
\label{Li_2} R_{ijk}\gtrsim R_{ijl},~~\text{if}~~ m_k\le m_l
\end{equation}

Note, however, that if $i,j$ are the light quarks, and the quarks
$k$ and $l$ are the heavy, then $R_{ijk}\approx R_{ijl}$ ({\it
e.g.,} $R_{qqc}\approx R_{qqb}$) in agreement with the limit of
the heavy quark effective theory.

Our estimations for the ratios $R_{ijk}/R_{ij}$ agree with the
results obtained using the nonrelativistic quark model or the bag
model  \cite{V85}-\cite{B94} or QCD sum rules \cite{CF96} which
are typically in the range $0.1-0.5$. On the other hand, our
result for $\Lambda_b$ disagrees with the one by Rosner \cite{R96}
who estimated the heavy-light diquark density at zero separation
in $\Lambda_b$ from the ratio of hyperfine splittings between
$\Sigma_b$ and
 $\Sigma_b^*$ baryons and $B$ and $B^*$ mesons and found
$R_{qbu}/R_{\bar b d} \sim 0.9\pm 0.1$, if the baryon splitting is
taken to be $m_{\Sigma^*_b}^2-m_{\Sigma_b}^2 \sim
m_{\Sigma^*_c}^2-m_{\Sigma_c}^2= (0.384\pm 0.035)~\text{GeV}^2$,
or even to $R_{ubd}/R_{\bar b d}\sim 1.8\pm 0.5$, if the
surprisingly small and not confirmed yet DELPHI result
$m_{\Sigma^*_b}-m_{\Sigma_b}=(56\pm 16)~\text{MeV}$ is used.

From the results of Table 3 it follows that the correlation
between two quarks depends on the third one. Note also that the
wave function calculated in HRA shows the marginal diquark
clustering in the doubly heavy baryons. This is principally
kinematic effect related to the fact that in the HRA the
difference between the various $\bar{r}_{ij}$ in a baryon is due
to the factor $\sqrt{1/\mu_{ij}}$ which varies between
$\sqrt{2/m_i}$ for $m_i=m_j$ and $\sqrt{1/m_i}$ for $m_i\ll m_j$.

In Table 5 we compare the short-range correlation coefficients in
the doubly heavy baryons with those calculated in \cite{BDGNR94}
using the pair-wise quark interaction with the power-law
potential, and in \cite{GKLO} using the non-relativistic model
with the Buchm\"uller--Tye potential.

\section{Masses of doubly heavy baryons}

To calculate hadron masses we, as in \cite{FS91}, first
renormalise the string potential \begin{equation} \label{C_i}
V_{\text{string}}\to V_{\text{string}}+\sum\limits_iC_i,
\end{equation} where the constants $C_i$ take into account the residual
self-energy (RSE) of quarks. In principle, these constants can be
expressed in terms of the two scalar functions entering covariant
expansion of the bilocal cumulants of gluonic fields in the QCD
vacuum \cite{Si89,FS91}. In the present work we treat them
phenomenologically. To find $C_i$ in (\ref{C_i}) we assume, first,
that the spin splittings of hadrons with a given quark content
arise from the colour-magnetic interaction in QCD. Indeed, for the
ground state hadrons the hadron wave functions have no orbital
angular momentum, so tensor and spin-orbit forces do not
contribute. The second assumption is  that the colour-magnetic
interaction can be treated perturbatively \cite{RGG75,KRQ87}:
\begin{equation} \label{FB} \Delta
E_{\text{spin}}=\frac{16\pi\alpha_s}{9}\sum\limits_{i<j}
\frac{\bsl{s}_i\bsl{s}_j}{m_im_j}R_{ijk}. \end{equation}

Because the colour-magnetic interaction between two quarks goes
inversely as the product of their masses, the perturbative
approximation improves as the quark mass increases. However, this
approximation may not be good for the baryons containing light
quarks\footnote{Note that $1/m_im_j$ dependence in Eq. (\ref{FB}),
if treated literally in the EH method, results in a collapse both
in the pseudoscalar $q\bar q$ channel and the proton. That may be
a signal of the Nambu--Goldstone phenomenon.}. In what follows we
adjust the RSE constants $C_i$ to reproduce the centre-of-gravity
for baryons with a given flavour. To this end we consider the
spin-averaged masses, such as:
\begin{equation}
\frac{M_N+M_{\Delta}}{2}=1.085~\text{GeV},~~~\text{and}~~~
 \frac{M_{\Lambda}+M_{\Sigma}+2M_{\Sigma^*}}{4}=1.267~\text{GeV},
\end{equation}
and analogous combinations for $qqc$ and $qqb$ states. Then we obtain
\begin{equation}
C_q=0.34~\text{GeV},~~~C_s=0.19~\text{GeV},~~~C_c\sim C_b\sim 0.
\end{equation}
We keep these parameters fixed to calculate the masses given in
Table 6, namely the spin-averaged masses (computed without the
spin-spin term) of the lowest doubly heavy baryons. Our results
are very similar to those obtained in \cite{BDGNR94} using the
pair-wise power-law potential.

\section{Conclusions}

In this paper we employ the general formalism for the baryons,
which is based on nonperturbative QCD and where the only inputs
are the string tension $\sigma$, the strong coupling constant
$\alpha_s$, and two additive constants, $C_q$ and $C_s$, the
residual self-energies of the light quarks. We present some
piloting calculations of the dynamical quark masses for various
baryons (see Table 1). The latters are computed solely in terms of
$\sigma$ and $\alpha_s$ and depend on a baryon.

The second important point of our investigation is the calculation
of the correlation functions for baryons. They are given, among
the other things, in Table 3. We have also performed the
calculations of the spin-averaged masses of baryons with two heavy
quarks. One can see from Table 6 that our predictions are
especially close to those obtained in \cite{BDGNR94} using a
variant of the power-law potential adjusted to fit ground state
baryons.

Evaluation of the spin-spin interactions requires inclusion of the
$K=2$ hyperspherical components and/or more sophisticated
treatment of the colour-magnetic interaction. We shall consider
these calculations in the next publication.

\section*{Acknowledgements}

We thank Yu.S.Kalashnikova and Yu. A. Simonov for useful
discussions. We also thank K.A.Ter-Martirosian  for his interest
in this work. This work was supported in part by RFBR grants \#\#
00-02-16363 and 00-15-96786.

\section*{Appendix}

\setcounter{equation}{0}
\def\theequation{A.\arabic{equation}}

Consider three quarks with arbitrary masses $m_i$, $i=1,2,3$, and
coordinates $\bsl{r}_i$. The problem is conveniently treated using
Jacobi coordinates $\bsl{\rho}_{ij}$ and $\bsl{\lambda}_{ij}$:
\begin{equation}
\label{jac}
\bsl{\rho}_{ij}=\alpha_{ij}(\bsl{r}_i-\bsl{r}_j),~~\bsl{\lambda}_{ij}=\beta_{ij}
\left(\frac{m_i\bsl{r}_i+m_j\bsl{r}_j}{m_i+m_j}-\bsl{r}_k\right),
\end{equation}
where
\begin{equation}
\label{alpha} \alpha_{ij}=\sqrt{\frac{\mu_{ij}}{\mu}},~~
\beta_{ij}=\sqrt{\frac{\mu_{ij,k}}{\mu}}.
\end{equation}
Here
$\mu_{ij}$ and $\mu_{ij,k}$ are the reduced masses
\begin{equation}
\mu_{ij}=\frac{m_im_j}{m_i+m_j},~~~
\mu_{ij,k}=\frac{(m_i+m_j)m_k}{m_i+m_j+m_k}\, .
\end{equation}
Altogether with the centre-of-mass coordinate $\rcm$ Jacobi
coordinates determine completely the position of the system. The
Jacobian of the transformation for the differential volume
elements is 1, {\it i.e.,} \begin{equation}
d^3\bsl{\rho}_{12}d^3\bsl{\lambda}_{12}=
d^3\bsl{\rho}_{32}d^3\bsl{\lambda}_{32}=
d^3\bsl{\rho}_{13}d^3\bsl{\lambda}_{13}.
\end{equation} The inverse transformations for the relative
coordinates $\bsl{r}_{ij}=\bsl{r}_i-\bsl{r}_j$ and
$\bsl{r}_k-\rcm$ are
\begin{equation} \label{inverse}
\bsl{r}_{ij}=\frac{1}{\alpha_{ij}}\bsl{\rho}_{ij},~~~
\bsl{r}_k-\rcm~
=~-\sqrt{\frac{\mu(m_i+m_j)}{m_k(m_1+m_2+m_3)}}~\bsl{\lambda}_{ij}\,
.
\end{equation}
The hyperradius $R^2$ is defined as
$R^2=\bsl{\rho}_{ij}^2+\bsl{\lambda}_{ij}^2$ and does not depend
on the order of the quark numbering:
\begin{equation} \label{r2}
R^2=\bsl{\rho}_{12}^2+\bsl{\lambda}_{12}^2=\bsl{\rho}_{32}^2+
\bsl{\lambda}_{32}^2=\bsl{\rho}_{13}^2+\bsl{\lambda}_{13}^2.
\end{equation}
Written in terms of $\bsl{r}_{ij}$ Eq. (\ref{r2}) reads:
\begin{equation}
R^2=\sum\limits_{i<j}\frac{m_im_j}{\mu(m_1+m_2+m_3)}\bsl{r}_{ij}^2\,
.
\end{equation}
In the centre-of-mass frame $\rcm=0$ the invariant
kinetic energy operator (\ref{H_0}) is written in terms the Jacobi
coordinates (\ref{jac}) as \begin{equation} \label{H_0_jacobi}
H_0= -\frac{1}{2\mu} \left(\frac{\partial^2}{\partial\bsl{\rho}^2}
+\frac{\partial^2}{\partial\bsl{\lambda}^2}\right)
=-\frac{1}{2\mu}\left( \frac{\partial^2}{\partial
R^2}+\frac{5}{R}\frac{\partial}{\partial R}+
\frac{K^2(\Omega)}{R^2}\right), \end{equation}
where $K^2(\Omega)$
is angular momentum operator whose eigen functions (the
hyperspherical harmonics) are
\begin{equation}
\label{eigenfunctions} K^2(\Omega)Y_{[K]}=-K(K+4)Y_{[K]},
\end{equation}
with $K$ being the grand orbital momentum. In terms
of $Y_{[K]}$ the wave function $\psi(\bsl{\rho},\bsl{\lambda})$
can be written in a symbolical shorthand as
$$\psi(\bsl{\rho},\bsl{\lambda})=\sum\limits_K\psi_K(R)Y_{[K]}(\Omega).$$
In the HRA $K=0$ and $\psi=\psi(R)$. Note that the centrifugal
potential in the Schr\"odinger equation for the radial function
$\psi_K(R)$ with a given $K$ $$ \frac{(K+2)^2-1/4}{R^2}$$ is not
zero even for $K=0$. For the reduced function
$\chi(R)=R^{5/2}\psi(R)$ one obtains after averaging the
interaction over the six-dimensional sphere Eq. (\ref{shr}) with
\begin{equation}
W(R)= \langle V(\bsl{\rho},\bsl{\lambda})\rangle =\int
~(V_c+V_{\text{string}})~\frac{d\Omega}{\pi^3}
\end{equation}

One can easily see that the definition of $\langle
V(\bsl{\rho},\bsl{\lambda})\rangle$ does not depend on the order
of the quark numeration.

In terms of the Jacobi coordinates the Coulomb and string
potentials read:
\begin{equation}
V_c=-\frac{2}{3}\alpha_s\sum\limits_{i<j}
\frac{\alpha_{ij}}{|\bsl{\rho}_{ij}|}, \end{equation}
\begin{equation} V_{\text{string}}=\sigma\sum\limits_{i<j}
\gamma_{ij} |\bsl{\lambda}_{ij}|, \end{equation} where
\begin{equation}
\label{gamma}
\gamma_{ij}=\sqrt{\frac{\mu(m_i+m_j)}{m_k(m_1+m_2+m_3)}}.
\end{equation}
Using the relations \cite{KNS87}
$$\langle\frac{1}{|\bsl{\rho}_{ij}|}\rangle=\frac{16}{3\pi}\cdot\frac{1}{R},~~~
\langle|\bsl{\lambda}_{ij}|\rangle=\frac{32}{15\pi}\cdot R, $$
valid for any pair (ij), one obtains Eqs. (\ref{ab}). \\[1mm]

\newpage

{\bfseries Table 1.} The constituent quark masses $m_i$ and the
ground state eigen energies $E_0$ (in units of GeV) for the various
baryon states. (The results obtained from the quasiclassical
solution and from the variational one practically coincide.)

\begin{center}
\begin{tabular}{|c|c|c|c|c|}
\hline baryon & $m_1$ & $m_2$ & $m_3$ & $E_0$ \\ \hline $(qqq)$ &
0.446 & 0.446 & 0.446 & 1.438 \\ \hline $(qqs)$ & 0.451 & 0.451 &
0.485 & 1.414
\\ \hline $(qss)$ & 0.457 & 0.490 & 0.490 & 1.392 \\ \hline $(sss)$ & 0.495 &
0.495 & 0.495 & 1.370 \\ \hline $(qqc)$ & 0.519 & 0.519 & 1.502 &
1.176 \\ \hline $(qsc)$ & 0.522 & 0.555 & 1.505 & 1.157 \\ \hline
$(ssc)$ & 0.589 & 0.589 & 1.507 & 1.138 \\ \hline $(qqb)$ & 0.564 &
0.564 & 4.836 & 1.057 \\ \hline $(qsb)$ & 0.567 & 0.601 & 4.837 &
1.038 \\ \hline $(ssb)$ & 0.604 & 0.604 & 4.838 & 1.019 \\ \hline
$(qcc)$ & 0.569 & 1.555 & 1.555 & 0.926 \\ \hline $(scc)$ & 0.604 &
1.557 & 1.557 & 0.908 \\ \hline $(qcb)$ & 0.606 & 1.616 & 4.866 &
0.783 \\ \hline $(scb)$ & 0.642 & 1.618 & 4.867 & 0.765 \\ \hline
$(qbb)$ & 0.636 & 4.931 & 4.931 & 0.582 \\ \hline $(sbb)$ & 0.673 &
4.931 & 4.931 & 0.565 \\ \hline
\end{tabular}
\end{center}

\newpage

{\bfseries Table 2.} The dynamical quark masses for the ground
state $(qc)$, $(sc)$, $(qb)$, $(sb)$ mesons \cite{KN00} and for the
corresponding ground state baryons.

\begin{center}
\begin{tabular}{|c|c|c|c|c|}
\hline
State & $m_q$ & $m_s$ & $m_c$ & $m_b$\\
\hline
$(qc)$ & 0.529 & & 1.497 &\\
$(sc)$ & & 0.569 & 1.501 & \\
$(qqc)$ & 0.519 & & 1.502 &\\
$(qsc)$ & 0.522 & 0.555 & 1.505 & \\
\hline
$(qb)$ & 0.619 &  &  & 4.84\\
$(sb)$ & &0.658  &  & 4.842\\
$(qqb)$ & 0.564 &  &  & 4.836\\
$(qsb)$ & 0.567 & 0.601  &  & 4.838\\
\hline
\end{tabular}
\end{center}

\newpage

{\bfseries Table 3.}~~$R_{ijk}$ in units of GeV$^3$ and
${\bar{r}_{ij}}=\sqrt{\langle\bsl{r}^2_{ij}\rangle}$ in units of
fm. (The results are obtained from the trial functions
(\ref{trial}) with the variational parameters $p_0$ given in units
of GeV${}^{(1/2)}$ in the first column. The results for light
baryons are presented for completeness.)

\begin{center}
\begin{tabular}{|c|c|c|c|c|c|c|c|}
\hline  baryon & $p_0$ & $R_{123}$ & $R_{231}$ & $R_{312}$ &
$\bar{r}_{12}$ & $\bar{r}_{23}$ & $\bar{r}_{31}$
\\ \hline $(qqq)$ & 0.472 & 0.00564 & 0.00564 & 0.00564 & 0.777 & 0.777 & 0.777
\\ \hline $(qqs)$ & 0.470 & 0.00567 & 0.00598 & 0.00598 & 0.775 & 0.762 & 0.762
\\ \hline $(qss)$ & 0.469 & 0.00600 & 0.00633 & 0.00600 & 0.760 & 0.747 & 0.760
\\ \hline $(sss)$ & 0.467 & 0.00636 & 0.00636 & 0.00636 & 0.746 & 0.746 & 0.746
\\ \hline $(qqc)$ & 0.454 & 0.00626 & 0.0113 & 0.0113 & 0.750 & 0.615 & 0.615
\\ \hline $(qsc)$ & 0.452 & 0.00656 & 0.0121 & 0.0113 & 0.738 & 0.601 & 0.615
\\ \hline $(ssc)$ & 0.451 & 0.00688 & 0.0121 & 0.0121 & 0.727 & 0.602 & 0.602
\\ \hline $(qqb)$ & 0.447 & 0.00681 & 0.0163 & 0.0163 & 0.729 & 0.545 & 0.545
\\ \hline $(qsb)$ & 0.446 & 0.00711 & 0.0176 & 0.0163 & 0.719 & 0.531 & 0.545
\\ \hline $(ssb)$ & 0.445 & 0.00742 & 0.0176 & 0.0176 & 0.708 & 0.531 & 0.531
\\ \hline $(qcc)$ & 0.439 & 0.0116 & 0.0296 & 0.0116 & 0.611 & 0.447 & 0.611
\\ \hline $(scc)$ & 0.438 & 0.0123 & 0.0294 & 0.0123 & 0.599 & 0.448 & 0.599
\\ \hline $(qcb)$ & 0.436 & 0.0123 & 0.0562 & 0.0166 & 0.599 & 0.361 & 0.541
\\ \hline $(scb)$ & 0.435 & 0.0130 & 0.0559 & 0.0178 & 0.587 & 0.361 & 0.529
\\ \hline $(qbb)$ & 0.438 & 0.0181 & 0.165 & 0.0181 & 0.527 & 0.252 & 0.527
\\ \hline $(sbb)$ & 0.437 & 0.0194 & 0.165 & 0.0194 & 0.515 & 0.252 & 0.515
\\ \hline
\end{tabular}
\end{center}

\vspace*{10mm}

{\bfseries Table 4.}~ The ratios of the squares of the wave
functions determining the probability to find a light quark at the
location of the heavy quark inside the heavy baryon and the
corresponding meson. (The meson wave functions are taken from
\cite{KN00}.)

\begin{center}
\begin{tabular}{|c|c|c|c|}
\hline $R_{ucd}/R_{uc} $ & $R_{scu}/R_{sc} $ &$R_{ubd}/R_{\bar bd}
$& $R_{sbu}/R_{sb}$ \\ \hline 0.436 & 0.405 & 0.373 & 0.340\\
\hline
\end{tabular}
\end{center}

\newpage

{\bfseries Table 5.} Short-range correlation coefficients
$R_{ijk}$. In the parentheses are shown the corresponding
quantities calculated using the power-law potential
\cite{BDGNR94}. In the square brackets are shown correlation
coefficients calculated using non-relativistic model with
Buchm\"uller-Tye potential.

\begin{center}
\begin{tabular}{|c|c|c|c|}
\hline State & $R_{123}$ & $R_{231}$ & $R_{312}$\\
\hline $(ccq)$
& 0.030~(0.039)~[0.022]  & 0.012~(0.009) & 0.012~(0.009)\\ $(ccs)$
& 0.030~(0.042)~[0.022]  & 0.012~(0.019) & 0.012~(0.019)\\ $(bbq)$
& 0.165~(0.152)~[0.144]  & 0.018~(0.012) & 0.018~(0.012)\\ $(bbs)$
& 0.165~(0.162)~[0.144]  & 0.019~(0.028) & 0.019~(0.028)\\ $(bcq)$
& 0.056~(0.065)~[0.042]  & 0.012~(0.010) & 0.017~(0.011)\\ $(bcs)$
& 0.056~(0.071)~[0.042]  & 0.013~(0.021) & 0.018~(0.025)\\ \hline
\end{tabular}
\end{center}

\vspace*{10mm}

{\bfseries Table 6.} Masses of baryons containing two heavy quarks

\begin{center}
\begin{tabular}{|c|c|c|c|c|c|c|}
\hline State & present & \cite{BDGNR94}$^{(a)}$ & \cite{E97}$^{(b)}$ &
\cite{LRP95}$^{(c)}$ & \cite{KKP94}
& \cite{LiOn}$^{(d)}$ \\ & work &&&&& \\ \hline $\Xi\{qcc\}$&
3.69 & 3.70 & 3.71 & 3.66 & 3.61 & 3.48\\ $\Omega\{scc\}$ & 3.86 &
3.80 & 3.76 & 3.74 & 3.71 & 3.58\\ \hline $\Xi\{qcb\}$    & 6.96 &
6.99 & 6.95 & 7.04 & & 6.82      \\ $\Omega\{scb\}$  & 7.13 & 7.07
& 7.05 & 7.09 & & 6.92     \\ \hline $\Xi\{qbb\}$    & 10.16 &
10.24 & 10.23 & 10.24 &  & 10.09
\\ $\Omega\{sbb\}$   & 10.34 & 10.30 & 10.32 & 10.37 &   & 10.19
\\ \hline
\end{tabular}
\end{center}

$^{(a)}$ The additive nonrelativistic quark model with the
power-law potential.\\ $^{(b)}$ Relativistic quasipotential quark
model.\\ $^{(c)}$ The Feynman-Hellmann theorem.\\ $^{(d)}$
Approximation of doubly heavy diquark.

\end{document}